# Papanicolaou Stain Unmixing for RGB Image Using Weighted Nucleus Sparsity and Total Variation Regularization

NANXIN GONG,[1,*] SAORI TAKEYAMA,[1] MASAHIRO YAMAGUCHI,[1] TAKUMI URATA,[1] FUMIKAZU KIMURA,[2] AND KEIKO ISHII[3]

[1] *Department of Information and Communications Engineering, Institute of Science Tokyo, Yokohama, Kanagawa, Japan*
[2] *Department of Biomedical Laboratory Sciences, Shinshu University, Asahi, Matsumoto, Nagano, Japan*
[3] *Division of Diagnostic Pathology, Okaya City Hospital, Honcho, Okaya, Nagano, Japan*

**Abstract**

The Papanicolaou stain, consisting of five dyes, provides extensive color information essential for cervical cancer cytological screening. The visual observation of these colors is subjective and difficult to characterize. Direct RGB quantification is unreliable because RGB intensities vary with staining and imaging conditions. Stain unmixing offers a promising alternative by quantifying dye amounts. In previous work, multispectral imaging was utilized to estimate the dye amounts of Papanicolaou stain. However, its application to RGB images presents a challenge since the number of dyes exceeds the three RGB channels. This paper proposes a novel training-free Papanicolaou stain unmixing method for RGB images. This model enforces (i) nonnegativity, (ii) weighted nucleus sparsity for hematoxylin, and (iii) total variation smoothness, resulting in a convex optimization problem. Our method achieved excellent performance in stain quantification when validated against the results of multispectral imaging. We further used it to distinguish cells in lobular endocervical glandular hyperplasia (LEGH), a precancerous gastric-type adenocarcinoma lesion, from normal endocervical cells. Stain abundance features clearly separated the two groups, and a classifier based on stain abundance achieved 98.0% accuracy. By converting subjective color impressions into numerical markers, this technique highlights the strong promise of RGB-based stain unmixing for quantitative diagnosis.

## 1 Introduction

Histopathology, involving the microscopic examination of tissue sections, is crucial for definitive cancer diagnosis, grading, and staging. Cytopathology, focusing on the analysis of exfoliated cells, is widely used for minimally invasive diagnostic procedures [1]. Cytopathology plays a key role in early cancer detection (e.g., cervical cancer screening). Through the Pap smear, cervical cells are collected and stained with the Papanicolaou stain, which discriminates cellular components by staining nuclei with hematoxylin (H), and cytoplasm with Eosin Y (EY), Light Green SF yellowish (LG), Orange G (OG) and Bismarck brown Y (BY), with intensity depending on factors like molecular weight, diffusivity, and differentiation.

Although the color in Papanicolaou stain is important in differentiating nuclei and cytoplasm of different types of cells, its visual characterization is often subjective and challenging to articulate. The application of digital image analysis is a potential solution for this purpose, and several methods have been proposed to quantify the colors of Papanicolaou-stained cytological specimen images. For instance, Nunobiki et al. introduced quantitative and qualitative analyses of stain color in RGB color spaces [2,3], while Sakamoto et al. quantified variations in cytoplasmic colors in the CIELAB color space [4]. However, the color in digital pathology imaging is affected by variations in staining processes and imaging devices [5]. Although color normalization and color augmentation techniques have been proposed, quantification based on physical quantities directly linked to the staining properties offers a distinct advantage over psychophysical quantities based on visual appearance.

* Corresponding author. gong.n.aa@m.titech.ac.jp (Nanxin Gong).

As a method to quantify colors using a physical quantity, estimating the stain abundance or dye amount based on spectral imaging has been proposed [6,7]. As for the application to the Papanicolaou stain, a dye amount estimation method based on 14-band multispectral (MS) images capturing Papanicolaou-stained specimens was developed [8]. This method employed MS images to estimate the dye amount in each pixel of cytological specimens via the pseudoinverse matrix that consists of the spectral absorption coefficients. However, MS cameras are not common in pathological imaging and require a lengthy scanning process. Therefore, it is expected to utilize RGB images in practice, which can be quickly scanned through a whole slide image (WSI) scanner [9].

Ruifrok et al. introduced a color deconvolution (CD) method to unmix RGB images into up to three dyes in converted optical density (OD) space, utilizing the OD vectors of pure dyes [10]. This approach assumes that the absorbance at each pixel is a linear combination of these dyes, and it computes the abundance of each dye by multiplying the observed OD values with the inverse of the stain OD matrix. Other traditional stain separation methods include independent component analysis (ICA) [11], singular value decomposition (SVD) [12], blind color deconvolution (BCD) [13], adaptive color deconvolution (ACD) [14], Bayesian K-SVD [15], stain color adaptive normalization (SCAN) [16], etc. These approaches typically focus on estimating the stain matrix, while the stain abundance is often computed using conventional CD. Most of these methods have been applied to H&E staining, where the number of dyes does not exceed the three RGB channels. However, when more than three dyes are involved, such as the Papanicolaou stain, the linear system becomes ill-posed.

Other approaches attempt to introduce constraints during the stain unmixing process to achieve a more accurate estimation. Nonnegative matrix factorization (NMF) has been adopted in an unsupervised stain unmixing framework, where both the stain color vectors and the corresponding stain abundances are simultaneously estimated for each image [17]. Vahadane et al. enhanced NMF by introducing a sparsity constraint, extending it to sparse NMF (SNMF), which preserves biological structure information during stain unmixing [18]. Nevertheless, when the number of dyes exceeds the available image channels, obtaining suitable stain vectors becomes difficult, often resulting in multiple valid but distinct solutions. Addressing the problem typically requires additional prior knowledge to mitigate ambiguities [19]. A group sparsity model-based algorithm has been proposed for unmixing more than three dyes in brightfield multiplex immunohistochemistry (IHC) images, but its applicability is limited to specific dyes due to reliance on biomarker co-localization [20].

A similar idea has also been explored in the field of spectral unmixing within remote sensing. Spectral unmixing aims to identify the reference spectra, quantify their spectral signatures, and compute their fractional abundances using observed mixed hyperspectral vectors [21]. The Sparse Unmixing via Variable Splitting and Augmented Lagrangian (SUnSAL) algorithm evaluates sparsity in the spectral domain using the l1 norm of the abundance matrix [22]. SUnSAL-TV further enhances this model by integrating total variation (TV) regularization, which promotes smoothness across neighboring pixels [23]. Although SUnSAL-TV was originally developed for spectral unmixing in remote sensing, its applicability to biomedical imaging has been previously explored [24].

Recently, deep learning approaches for stain unmixing have emerged. Duggal et al. proposed a stain deconvolutional layer (SD-layer) that can be prefixed to any Convolutional Neural Network (CNN) model [25]. Zheng et al. introduced a capsule network-based stain standardization module that can learn and generate uniform stain separation outputs [26]. Chen et al. developed a physics-guided deep image prior network (PGDIPS) for stain deconvolution, leveraging a self-supervised deep neural network to separate an image into an arbitrary number of dyes [27]. Abousamra et al. proposed autoencoder-based approaches (ColorAE [28] and InverseAE [29]) for multiplex brightfield immunohistochemical (IHC) stain decomposition and compared them with U-Net [30]. Their methods focused on biomarker segmentation rather than quantitative stain unmixing. Ghahremani et al. introduced DeepLIIF, a GAN-based multi-task deep learning framework that performs stain deconvolution [31]. Yang et al. proposed a variational Bayesian blind CD network (BCD-Net), which is promising for H&E [32]. However, all these approaches are generally limited in that they (i) are applicable only when the number of dyes does not exceed three, such as H&E staining, (ii) primarily focus on segmenting different stained regions rather than providing quantitative abundances, and/or (iii) rely on supervised training with extensive ground truth data.

In summary, the existing methods have not adequately addressed stain unmixing in scenarios where the number of dyes exceeds the number of image channels and where there is extensive overlap among dyes. The sparsity assumption, common in stain separation and hyperspectral unmixing, is not applicable to Papanicolaou-stained images, as nearly all dyes are typically present across the image, with multiple dyes frequently coexisting within the same pixel. This limitation has motivated the design of a new stain unmixing method for Papanicolaou staining. In this study, we propose a training-free stain unmixing method tailored to Papanicolaou-stained RGB images, which is optimization-based and incorporates prior knowledge of stain abundance characteristics. Specifically, the method leverages the following three properties:

- Nonnegativity: At any given pixel, stain abundance and its corresponding OD value must be nonnegative. Negative values would suggest an unrealistic emission of light.
- Nucleus sparsity: In Papanicolaou-stained samples, H typically binds to nuclei, whereas the cytoplasm is nearly unstained by H. To reflect this distribution, we introduce a sparsity-promoting regularization term that iteratively assigns higher weights to pixels with lower H abundance and lower weights to higher H abundance, thereby enhancing sparsity in non-nuclear regions while preserving signals in nuclei.
- Piecewise smoothness: Due to the continuity of cellular structures, stain abundances are expected to vary smoothly within homogeneous regions. We model this property using TV regularization to encourage spatial coherence among neighboring pixels.

In stain unmixing experiments, obtaining ground truth for stain matrix and stain abundances presents significant challenges. Traditional methods often rely on manually selecting representative stained regions, which can introduce subjective bias. To mitigate this, we prepared single-stain specimens to compute the stain matrix. Additionally, following the methodology described in [8], we utilized MS stain unmixing results as ground truth. Furthermore, we conducted benchmark comparisons between our proposed method and existing stain unmixing algorithms to evaluate performance in stain quantification.

To further validate the applicability of our stain unmixing method, we applied it to quantify the colors of the cytoplasmic mucin in normal endocervical (EC) cells and lobular endocervical glandular hyperplasia (LEGH) cells. LEGH is reported to be a benign glandular tumor of the uterine cervix [33]. As it is considered a precursor lesion of gastric-type endocervical adenocarcinoma, early detection and accurate diagnosis of LEGH are of great significance for a good prognosis. Despite its importance, LEGH cells lack clear nuclear atypia, and it is difficult to distinguish them from EC cells morphologically by human eyes [34,35]. According to Ishii et al., the gastric-type neutral mucin in the cytoplasm of LEGH cells exhibits yellowish, whereas the acidic mucin of EC cells shows pinkish when stained with Papanicolaou staining [36-38]. This color distinction aids in identifying LEGH cells. However, diagnosing LEGH remains subjective, lacking a quantitative criterion based on mucin color, underscoring the need for objective diagnostic measures.

## 2 Multispectral stain unmixing method

In this section, we introduce the method for performing Papanicolaou stain unmixing using MS images [8]. The stain unmixing results from MS images are considered the ground truth and serve as references for analyzing the Papanicolaou stain unmixing results from RGB images.

By assuming the Beer-Lambert law, for a certain wavelength λ, the absorbance (i.e., OD) of observation $\alpha(\lambda)$ is modeled by the following formulations:

$$\alpha(\lambda) = \textstyle\sum_i c_i \varepsilon_i(\lambda), \tag{1}$$

$$I_T(\lambda) = 10^{-\alpha(\lambda)} I_I(\lambda). \tag{2}$$

where $c_i$ is the dye amount and $\varepsilon_i(\lambda)$ is the spectral absorption coefficient of the $i$-th stain ($i = 1, \ldots, R$, and $R$ is the number of dyes). Let $I_I(\lambda)$ and $I_T(\lambda)$ be the intensities of the incident and transmitted light, which are obtained from the glass and stained pixels, respectively. According to Eq. (2) the spectral absorbance $\alpha(\lambda)$ can be calculated by

$$\alpha(\lambda) = \log_{10}\left(\frac{I_I(\lambda)}{I_T(\lambda)}\right). \tag{3}$$

We can define the spectral absorbance of the $k$-th spectral channel $\alpha(\lambda_k)$ as

$$\alpha(\lambda_k) = \log_{10}\left(\frac{I_I(\lambda_k)}{I_T(\lambda_k)}\right). \tag{4}$$

Then, for each pixel, we get

$$\begin{gathered}
\mathbf{y} = \mathbf{E}\mathbf{x}, \\
\text{s.t. } \mathbf{y} = [\alpha(\lambda_1), \alpha(\lambda_2), \ldots, \alpha(\lambda_M)]^{\mathrm{T}} \in \mathbb{R}^{M\times 1}, \\
\mathbf{E} = \begin{bmatrix} \varepsilon_1(\lambda_1) & \varepsilon_2(\lambda_1) & \cdots & \varepsilon_R(\lambda_1) \\ \varepsilon_1(\lambda_2) & \varepsilon_2(\lambda_2) & & \varepsilon_R(\lambda_2) \\ \vdots & & \ddots & \vdots \\ \varepsilon_1(\lambda_M) & \varepsilon_2(\lambda_M) & \cdots & \varepsilon_R(\lambda_M) \end{bmatrix} \in \mathbb{R}^{M\times R}, \\
\mathbf{x} = [c_1, c_2, \ldots, c_R]^{\mathrm{T}} \in \mathbb{R}^{R\times 1}.
\end{gathered} \tag{5}$$

where $M$ is the number of spectral channels. The spectral absorption coefficient matrix $\mathbf{E}$ is normalized such that each column has a unit summation of its elements:

$$\mathbf{1}^{\mathrm{T}}\mathbf{e}_i = 1, \tag{6}$$

where $\mathbf{1}$ is an $M$-dimensional column vector of ones, and $\mathbf{e}_i$ is the $i$-th column vector of $\mathbf{E}$. In [8], 14-band MS images with spectral information from 440 nm to 720 nm at 20 nm intervals were captured. Since $R < M$ in MS Papanicolaou stain unmixing, the stain abundance $\mathbf{x}$ of the pixel can be estimated by multiplying the pseudoinverse of the spectral absorption coefficient matrix $\mathbf{E}$ with the absorbance vector $\mathbf{y}$ of the captured MS image. The stain absorption matrix $\mathbf{E}$ was computed in advance based on Eq. (4) using MS images of single-stain specimens. Please refer to [8] for more details on MS stain unmixing.

## 3 Proposed RGB stain unmixing method

In this section, we present the methodology for the proposed stain unmixing algorithm using RGB images. Fig. 1 illustrates the pipeline for Papanicolaou stain unmixing, which includes RGB image acquisition, OD conversion, stain unmixing with regularizations, and stain abundance estimation. To address the underdetermined nature of the problem, we leverage a stain matrix pre-measured from single-stain RGB images together with three regularizations related to stain abundances.

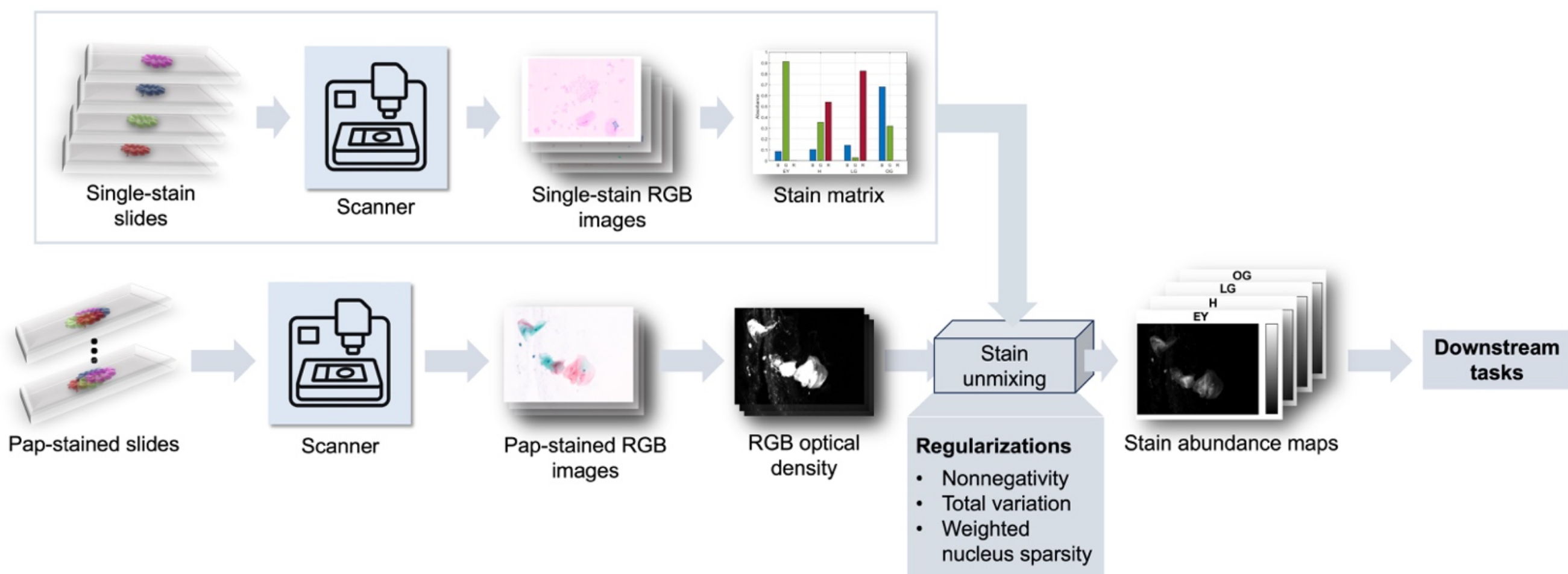

**Fig. 1** Pipeline for Papanicolaou stain unmixing using RGB images. Single-stain slides are first scanned to estimate the stain matrix. Papanicolaou-stained images are then scanned and converted into the OD space. The stain unmixing process decomposes each pixel into stain abundance maps corresponding to individual dyes (EY, H, LG, and OG) under the constraint of several regularizations, including nonnegativity, TV, and weighted nucleus sparsity. The resulting abundance maps can be further utilized for downstream tasks

### *3.1 Formulation of the RGB stain unmixing problem*

Let $\mathbf{I}_1 \in \mathbb{R}^{3\times n}$ be the matrix of observed RGB intensities in an image, where $n$ is the number of pixels, and let $\mathbf{I}_0 \in \mathbb{R}^{3\times 1}$ be the RGB (red, green, and blue) intensity of the incident light on the sample (obtained from the glass pixels). For the $i$-th spectral channel ($i = 1, 2, 3$), the absorbance (i.e., OD) $\mathbf{y}_i \in \mathbb{R}^{1\times n}$ is

$$\mathbf{y}_i = \log_{10} \frac{\boldsymbol{l}_i}{I_{0,i}}. \tag{7}$$

where $\boldsymbol{l}_i$ is the $i$-th row vector of $\mathbf{I}_1$, and $I_{0,i}$ is the $i$-th element of $\mathbf{I}_0$. The influence of the spectral sensitivities of RGB channels is neglected for simplicity.

Stained tissue attenuates light in a certain spectrum depending on the type and amount of dye it has absorbed. Let $\mathbf{A} \in \mathbb{R}^{3\times r}$ be the stain absorption coefficient matrix (i.e., stain matrix) whose columns represent the absorbance in RGB channels of each dye such that $r$ is the number of dyes, $\mathbf{X} \in \mathbb{R}^{r\times n}$ be the stain abundance matrix, where the row represents the amount of each dye, and $\mathbf{Y} \in \mathbb{R}^{3\times n}$ be the OD of RGB channels. Then, we can write

$$\mathbf{Y} = \mathbf{AX}. \tag{8}$$

When both $\mathbf{A}$ and $\mathbf{X}$ are unknown, it is called a blind unmixing problem, and several methods have been proposed [11-15]. Blind unmixing is often employed as an intermediate step in color normalization, where distinct stain matrices are estimated for images exhibiting color variations, and subsequently used in CD [10] to derive the stain abundance matrix. However, in our implementation the stain matrix is fixed across samples and derived from single-stain slides. In routine practice, slides processed under a consistent protocol within one laboratory (same day/batch with appropriate quality control) should share a common stain matrix; estimating a different stain matrix per image is unwarranted and may introduce unnecessary variability.

Therefore, we use single-stain RGB images to measure the stain matrix $\mathbf{A}$ in advance. We excluded BY from the study and considered only the effects of the remaining four dyes (EY, H, LG, and OG) since BY is almost transparent in our case, resulting in $r = 4$. After measuring the stain absorption coefficients of the individual dyes, we perform normalization to ensure that each column vector has a unit summation, ensuring that every dye has a comparable contribution to the calculation

$$\mathbf{1}^{\mathrm{T}}\mathbf{a}_i = 1, \tag{9}$$

where $\mathbf{1}$ is a three-dimensional column vector of ones, and $\mathbf{a}_i$ is the $i$-th column vector of $\mathbf{A}$.

As a result, estimating $\mathbf{X}$ becomes a linear inverse problem. However, a challenge arises because the number of dyes in the Papanicolaou stain exceeds the RGB image channels, making the unmixing problem underdetermined. The proposed method employs prior knowledge to estimate the stain abundance $\mathbf{X}$.

### 3.2 *Proposed unmixing method based on nonnegativity, weighted nucleus sparsity, and TV regularizations*

We convert the RGB image to OD space and estimate at every pixel how much of each dye is present. Because there are more dyes than color channels, we make the problem solvable by using the stain matrix measured from single-stain images and encoding three prior knowledge of stain abundance. Considering that in Papanicolaou-stained samples the stain abundance cannot be negative, we introduce a nonnegativity regularization. Additionally, we incorporate TV regularization to account for the piecewise smoothness of abundance in the neighboring pixels. Furthermore, H abundance is sparse in the image since H typically binds with the cell nuclei and the cytoplasm is rarely stained by H. We propose a novel regularization, weighted nucleus sparsity regularization, to enhance the sparsity of H abundance.

Based on the aforementioned constraints, we formulate the problem of Papanicolaou stain unmixing in RGB images into a regularized linear regression problem. We perform stain unmixing by solving the following optimization problem:

$$\min_{\mathbf{X}} \frac{1}{2}\|\mathbf{AX}-\mathbf{Y}\|_F^2 + \lambda\|\mathbf{w}\odot\mathbf{x}_H\|_{1,1} + \lambda_{TV} TV(\mathbf{X}) \ \mathrm{s.t.}: \mathbf{X} \geq 0, \tag{10}$$

where $\odot$ denotes the elementwise multiplication of two variables, $\mathbf{x}_H$ is a $1 \times n$ row vector representing the abundance of H, and

$$TV(\mathbf{X}) \equiv \sum_{\{i,j\}\in\varepsilon} \left\|\mathbf{x}_i - \mathbf{x}_j\right\|_1, \tag{11}$$

is a vector extension of the anisotropic TV [39,40], which promotes piecewise constant (or smooth) transitions in the fractional abundance of the same dye among neighboring pixels, $\varepsilon$ denotes the set of horizontal and vertical neighbors in the image, and $\mathbf{x}_i$ represents the $i$-th column of $\mathbf{X}$.

Let $\mathbf{H}_h: \mathbb{R}^{m\times n} \to \mathbb{R}^{m\times n}$ denote a linear operator computing the horizontal differences between the components of $\mathbf{X}$ corresponding to neighboring pixels; i.e., $\mathbf{H}_h\mathbf{X} = [\mathbf{d}_1, \mathbf{d}_2, \dots, \mathbf{d}_n]$, where $\mathbf{d}_i = \mathbf{x}_i - \mathbf{x}_{i_h}$ with $i$ and $i_h$ denoting a pixel and its horizontal neighbor. We are assuming periodic boundaries. Let $\mathbf{H}_v: \mathbb{R}^{m\times n} \to \mathbb{R}^{m\times n}$ be defined in a similar way for the vertical differences; i.e., $\mathbf{H}_v\mathbf{X} = [\mathbf{v}_1, \mathbf{v}_2, \dots, \mathbf{v}_n]$, where $\mathbf{v}_i = \mathbf{x}_i - \mathbf{x}_{i_v}$, with $i$ and $i_v$ denoting a pixel and its vertical neighbor. With these two difference operators, we define

$$\mathbf{HX} \equiv \begin{bmatrix} \mathbf{H}_h\mathbf{X} \\ \mathbf{H}_v\mathbf{X} \end{bmatrix}. \tag{12}$$

At iteration $t$, let the stain abundance matrix $\mathbf{X}^{(t)}$ consist of four $1 \times n$ row vectors $\mathbf{x}_s^{(t)}$ $(s \in \{EY, H, LG, OG\})$, each corresponding to the abundance of one dye (EY, H, LG, and OG).

$$\mathbf{X}^{(t)} = \begin{bmatrix} \mathbf{x}_{EY}^{(t)} \\ \mathbf{x}_{H}^{(t)} \\ \mathbf{x}_{LG}^{(t)} \\ \mathbf{x}_{OG}^{(t)} \end{bmatrix}. \tag{13}$$

Let $\mathbf{w}^{(t+1)}$ be the nucleus sparsity weight vector at iteration $t+1$ and it is calculated using the H abundance at iteration $t$, $\mathbf{x}_H^{(t)}$:

$$\mathbf{w}^{(t+1)} = \exp\left(-\mathbf{x}_H^{(t)}\right). \tag{14}$$

When updating the weights, we constrain the H abundance to be nonnegative ($x_{H,i} \geq 0$ for all $i$). Consequently, the resulting weights satisfy $\mathbf{w} \in (0, 1]^n$, which effectively prevents extreme values and maintains numerical stability throughout the optimization process. Then we can use

$$\mathbf{W}^{(t)} = \left[\mathbf{0}\ \mathbf{w}^{(t)\mathrm{T}}\ \mathbf{0}\ \mathbf{0}\right]^{\mathrm{T}}. \tag{15}$$

to represent the nucleus sparsity weight matrix of the same dimension as $\mathbf{X}$ at iteration $t$, obtained by padding the nucleus sparsity weight vector with an $n \times 1$ zero-vector $\mathbf{0}$.

With these definitions in place, Prob. (10) can be reformulated as follows:

$$\min_{\mathbf{X}} \frac{1}{2}\|\mathbf{AX}-\mathbf{Y}\|_F^2 + \lambda\|\mathbf{W}\odot\mathbf{X}\|_{1,1} + \lambda_{TV}\|\mathbf{HX}\|_{1,1} + \iota_{R+}(\mathbf{X}), \tag{16}$$

where $\iota_{R+}(\mathbf{X}) = \sum_{i=1}^{n} \iota_{R+}(\mathbf{x}_i)$ is the indicator function of the nonnegative real number set $\mathbb{R}^+$ ($\iota_{R+}(\mathbf{x}_i)$ is zero if $\mathbf{x}_i$ belongs to the nonnegative orthant and $+\infty$ otherwise). Prior knowledge that hematoxylin stains nuclei rather than cytoplasm is encoded by making the weight a decreasing function of the local H abundance: a lower H abundance in the cytoplasm is assigned a higher constraint weight in the subsequent iteration, effectively reducing the small abundance to zero. Conversely, a larger H abundance in the nucleus receives a lower weight, thus preserving the abundance.

To solve this optimization problem, we construct an algorithm based on the alternating direction method of multipliers (ADMM) [41]. The details of the optimization and computational complexity can be found in Online Resource 1. The implementation of our method is publicly available.

### *3.3 Abundance normalization by reference values*

When presenting quantitative results, normalizing the estimated stain abundance to the range $[0, 1]$ makes it easier to observe and interpret. To achieve this, we determined the reference values for each dye by selecting well-stained areas from the Papanicolaou stain images. Then, we considered the robust maximum value of stain abundance (e.g., the top 1% value) as the reference value $q_i$ $(i = 1,2,3,4)$. The normalized stain abundance was obtained by dividing the actual stain abundance by this reference value.

## 4 Experiments and results

This section begins by describing the preparation of Papanicolaou-stained and single-stain slides. We then explain how the RGB images for unmixing were obtained. Then, the estimation of the stain matrix is demonstrated. The proposed method was evaluated using RGB images converted from MS data, with the unmixing results from MS stain unmixing serving as the ground truth. To quantitatively evaluate our proposed RGB stain unmixing, we calculated the calibration coefficients between MS and RGB stain unmixing results using single-stain images. Additionally, we assessed the performance of our method in terms of stain quantification and cell sample classification.

### *4.1 Specimen preparation*

Six Papanicolaou-stained uterine cervix specimens were collected from six patients, including three EC cases and three LEGH cases. The staining was performed using Carrazzi's hematoxylin solution (Hematoxylin, Sigma-Aldrich, STL, MO, USA), OG-6 (Muto Pure Chemicals Co., Ltd.), and EA-50 (Muto Pure Chemicals Co., Ltd.). For the single-stain procedure, LG, EY, and BY were employed to replicate the composition found in EA-50, creating individual stain solutions. These solutions were then used to stain the collected uterine cervix specimens, producing single-stain results. The staining times were set as follows: 85 seconds for Carrazzi's hematoxylin, 65 seconds for OG-6, and 170 seconds for each dye within EA-50 (LG, EY, BY). For each dye, three single-stain uterine cervix specimens were prepared. All specimens were prepared simultaneously using standardized protocols and stored under identical environmental conditions to minimize potential color variations arising from differences in staining procedures or storage environments.

### *4.2 RGB image acquisition*

In the experiments, we employed the MS microscopic camera (Vectra®3, PerkinElmer Inc.) to capture the spectral information of specimens. Samples were captured at 40× magnification across 14 spectral bands ranging from 440 nm to 720 nm, with 20 nm intervals between bands. The resulting images had a resolution of 1020×1368 pixels. For subsequent experimental analysis, we converted MS images into RGB images. Since we aim to perform pixel-level comparisons between stain unmixing results from RGB and MS images, generating RGB images from MS data minimizes discrepancies arising from differences in imaging systems—such as spatial resolution, focus areas, depth of field, and image misalignment—ensuring that performance evaluation metrics reflect only the effects of the stain unmixing methods themselves.

To generate RGB images from spectral data of MS images, we first converted the spectral data into CIE XYZ values. We then derived linear RGB values from XYZ and, after gamma correction, obtained the final standard RGB (sRGB) image. Since Papanicolaou staining is a transmissive case, we used CIE standard illuminant D65 as the reference illuminant in conversion.

Fig. 2 compares two RGB images of the same region of interest (ROI) from a Papanicolaou-stained slide of a LEGH case: (a) MS-RGB, an sRGB image synthesized from the 14-band MS cube, and (b) WSI-RGB, a crop from a WSI acquired with a NanoZoomer 2.0-HT scanner (Hamamatsu Photonics) at 40× magnification. The two panels were rigidly registered and displayed at the same pixel size. Minor hue differences are expected: MS-RGB results from colorimetric rendering, whereas WSI-RGB reflects the scanner's proprietary spectral sensitivities and may include embedded processing. Because accurate color reproduction was not the focus of this study, in our experiment, we used MS-RGB as the RGB input. In the subsequent text, the RGB images refer to sRGB images converted from MS images, unless otherwise stated.

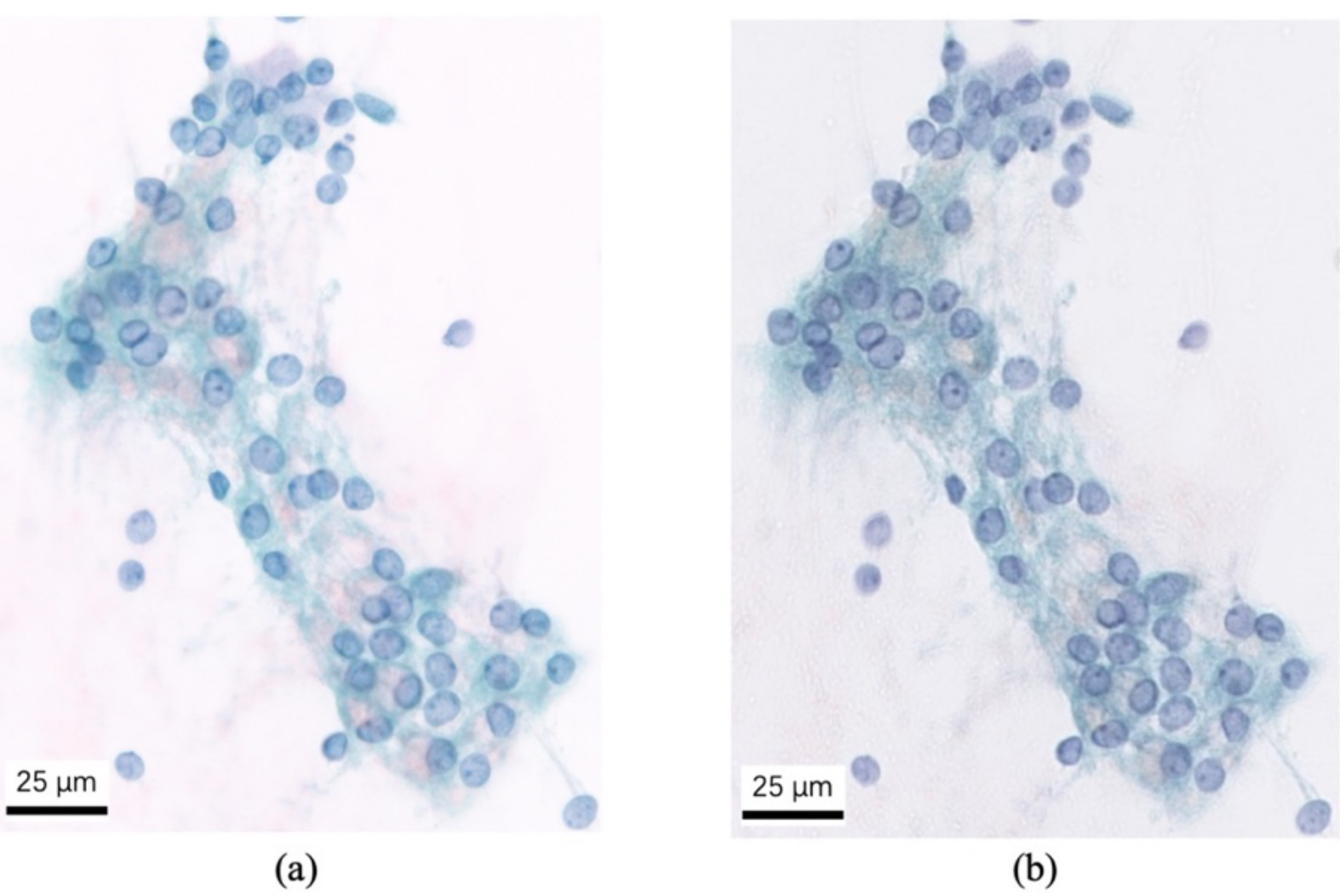

**Fig. 2** Papanicolaou-stained images of the same ROI shown as (a) MS-RGB (sRGB generated from MS cube) and (b) WSI-RGB (crop from NanoZoomer WSI)

As a result, we initially captured MS images from single-stain specimens for each dye. Corresponding RGB images were generated from these MS images (Table 1) to facilitate the measurement of stain matrix (see Section 0). Subsequently, we acquired MS images from several Papanicolaou-stained specimens and converted them into RGB images. Table 2 summarizes the number of images containing EC and LEGH cells captured from each Papanicolaou-stained slide, as well as the number of cropped 256×256 pixel patches. These images were utilized to evaluate the quantification performance of the proposed RGB stain unmixing method (see Section 4.5) and the cell classification performance (see Section 4.7). Four cases (including two EC and two LEGH) were used for parameter tuning or training, while the remaining two cases (including one EC and one LEGH) were reserved for testing.

**Table 1** The number of single-stain images

| | EY | H | LG | OG | BY |
|---|---|---|---|---|---|
| Number | 18 | 21 | 17 | 19 | 18 |

**Table 2** The number of Papanicolaou-stained images

| Slide No. (Annotation) | Slide 1 (EC) | Slide 2 (EC) | Slide 3 (EC) | Slide 4 (LEGH) | Slide 5 (LEGH) | Slide 6 (LEGH) |
|---|---|---|---|---|---|---|
| Number of images | 5 | 10 | 11 | 10 | 14 | 27 |
| Number of 256×256 patches | 76 | 184 | 160 | 96 | 94 | 316 |
| | For parameter tuning or training | | For testing | | For parameter tuning or training | |

Additionally, to examine the robustness of our stain unmixing method across different RGB images, we performed it on MS-RGB and WSI-RGB datasets, respectively. Resulting stain matrices and preliminary unmixing results are provided in Online Resource 2, which confirm the effectiveness of the method in adapting to different stain matrices and mitigating color variations introduced by imaging systems. A more systematic and comprehensive study on multi-center data will be conducted and reported in future work.

### 4.3 Measurement of stain matrix

We used single-stain images to determine the stain matrices required for unmixing. Some captured images are displayed in Fig. 3. In Papanicolaou staining, BY is a weak, often-omitted counterstain. In our dataset, typical BY staining was not observed on the Papanicolaou slides, and even on BY single-stain slides the coloration was very faint. The BY absorbance shows low contrast, making the five-dye unmixing problem ill-conditioned. Thus, we considered only EY, H, LG, and OG dyes in our study. As illustrated by the blue-shaded areas in Fig. 3, we cropped eight well-stained samples from the single-stain images for each dye, then calculated, averaged, and normalized their absorbances to get the absorption coefficients.

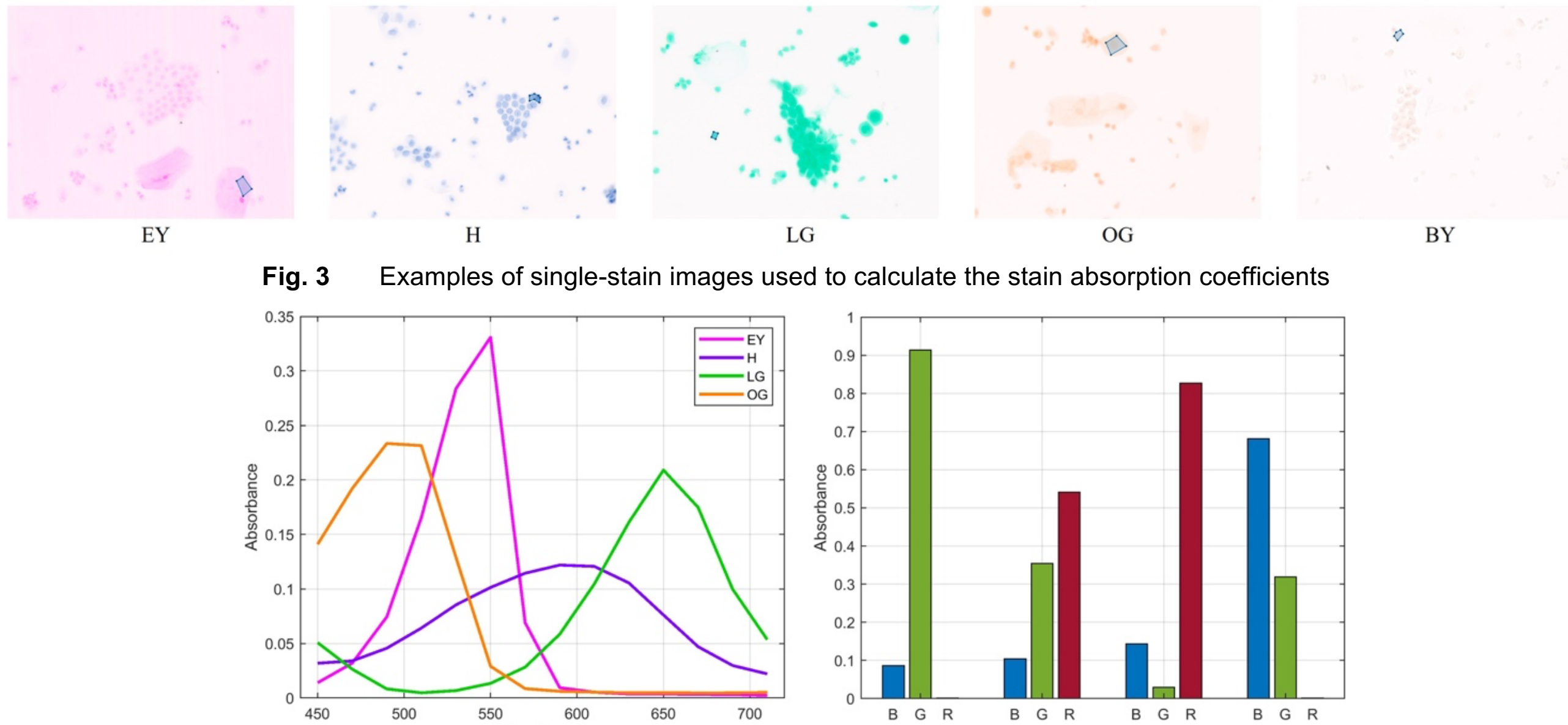

**Fig. 3** Examples of single-stain images used to calculate the stain absorption coefficients

**Fig. 4** Normalized MS absorption coefficients (left), and RGB absorption coefficients (right) of Papanicolaou stain

Fig. 4 illustrates the 14-band MS and RGB absorption coefficients of EY, H, LG, and OG. It can be observed that the conventional H and EY exhibit some spectral overlap within their primary absorption bands; however, the overall spectral differentiation remains relatively clear. With the inclusion of LG and OG, the spectral overlap among the dyes becomes more intricate. For instance, significant overlap between OG and EY occurs in the 500–530 nm range, while notable intersections between H and LG are found within the 600–660 nm range, and between H and OG around 520 nm. These overlaps impose greater challenges on color unmixing algorithms, necessitating additional constraints to achieve accurate stain unmixing.

### 4.4 Abundance calibration between MS and RGB stain unmixing

The results of MS stain unmixing were considered the ground truth and served as references for analyzing the results of RGB stain unmixing. To enable the quantitative comparison, we calculated the calibration coefficient, which is the ratio of the stain abundances obtained by MS stain unmixing to those obtained by RGB stain unmixing. We calculated a calibration coefficient for each dye.

For instance, with the $i$-th dye $(i = 1,2,3,4)$, we began by selecting a typical single-stain sample and calculating its MS absorbance $\mathbf{Y}_i \in \mathbb{R}^{14\times m}$ and RGB absorbance $\mathbf{Y}_i' \in \mathbb{R}^{3\times m}$ , where $m$ is the number of pixels in the sample. Here, we assume the presence of only that dye in the sample, so we have

$$\boldsymbol{e}_i \boldsymbol{s}_i = \mathbf{Y}_i, \tag{17}$$

$$\boldsymbol{a}_i \boldsymbol{s}_i' = \mathbf{Y}_i', \tag{18}$$

where $\boldsymbol{e}_i \in \mathbb{R}^{14\times 1}$ is the MS absorption coefficient vector of the $i$-th dye, $\boldsymbol{a}_i \in \mathbb{R}^{3\times 1}$ is the RGB absorption coefficient vector of this dye, $\boldsymbol{s}_i \in \mathbb{R}^{m}$ and $\boldsymbol{s}_i' \in \mathbb{R}^{m}$ denote the MS and RGB stain abundance vector of the single-stain sample, respectively. $\boldsymbol{s}_i$ and $\boldsymbol{s}_i'$ were computed using the least squares method.

Subsequently, we used least squares to calculate the calibration coefficient $p_i$:

$$p_i = \arg\min_{p_i} \sum_{j=1}^{m} (p_i \boldsymbol{s}_{ij}' - \boldsymbol{s}_{ij})^2 = \frac{\boldsymbol{s}_i' \cdot \boldsymbol{s}_i}{\boldsymbol{s}_i' \cdot \boldsymbol{s}_i'}. \tag{19}$$

The calibration coefficients between the stain abundances obtained by MS stain unmixing and those obtained by RGB stain unmixing were $p_1 = 7.51, p_2 = 13.13, p_3 = 12.36, p_4 = 5.79$. It is important to note that this calibration was intended solely to facilitate numerical comparisons with MS stain unmixing results; it is not a necessary step for performing RGB stain unmixing itself.

### *4.5 Evaluation of stain quantification performance*

We evaluated the proposed RGB stain unmixing algorithm using the generated RGB images of the Papanicolaou-stained specimens. As described in Section 3.3, we normalized all stain abundances within the range $[0, 1]$. The reference values for normalizing stain abundance were determined as $q_1 = 3.20, q_2 = 6.87, q_3 = 5.78, q_4 = 2.97$.

For quantitative analysis, the MS stain unmixing results were considered the ground truth. To evaluate the unmixing accuracy, we employed two widely recognized metrics: Signal-to-Reconstruction Error (SRE), measured in decibels (dB), and Root Mean Square Error (RMSE). SRE assesses the fidelity of the reconstructed signal relative to the original, with higher values indicating better reconstruction quality. RMSE quantifies the average magnitude of the error between the estimated and true values, where lower values signify more accurate estimations.

$$\mathrm{SRE(dB)} = 10\log_{10}(E(\|\mathbf{X}_{GT}\|_F^2)/E(\|\mathbf{X}_{GT} - \widehat{\mathbf{X}}\|_F^2)), \quad (20)$$

$$\mathrm{RMSE} = \sqrt{\|\mathbf{X}_{GT} - \widehat{\mathbf{X}}\|_F^2/n}, \quad (21)$$

where $\widehat{\mathbf{X}}$ is the abundance estimated by the RGB stain unmixing method, $\mathbf{X}_{GT}$ is the ground truth, $E(\cdot)$ denotes the expectation function, and $n$ is the total number of elements in the matrix (i.e., rows × columns × bands).

For the comparative analysis with the quantification results of the proposed method, we also report the results of traditional methods, namely CD [10], SNMF [18], and SUnSAL-TV [23], as well as deep learning-based methods, including ColorAE [28] and U-Net [30]. To the best of our knowledge, no standardized benchmark or widely accepted unmixing algorithm exists for Papanicolaou-stained images. Consequently, these methods were adapted for application to Papanicolaou stain unmixing. CD was originally developed for H&E staining, so we adapted it for Papanicolaou stain by computing the Moore-Penrose pseudoinverse of the stain matrix as the color deconvolution matrix. SNMF, being a blind stain separation technique that jointly estimates both the stain matrix and stain abundance matrix, was modified to estimate only the stain abundances using the fixed stain matrix to ensure a fair comparison. SUnSAL-TV was originally developed for spectral unmixing in remote sensing, whose applicability to biomedical imaging has been explored in [24]. It is an optimization-based method with nonnegativity, l1 sparsity, and TV regularization. ColorAE is an autoencoder-based model that combines a reconstruction loss with an abundance loss, which has been demonstrated to successfully separate more biomarkers than the number of imaging channels. U-Net introduces skip connections within the CNN architecture and has proven efficient in various medical image segmentation tasks.

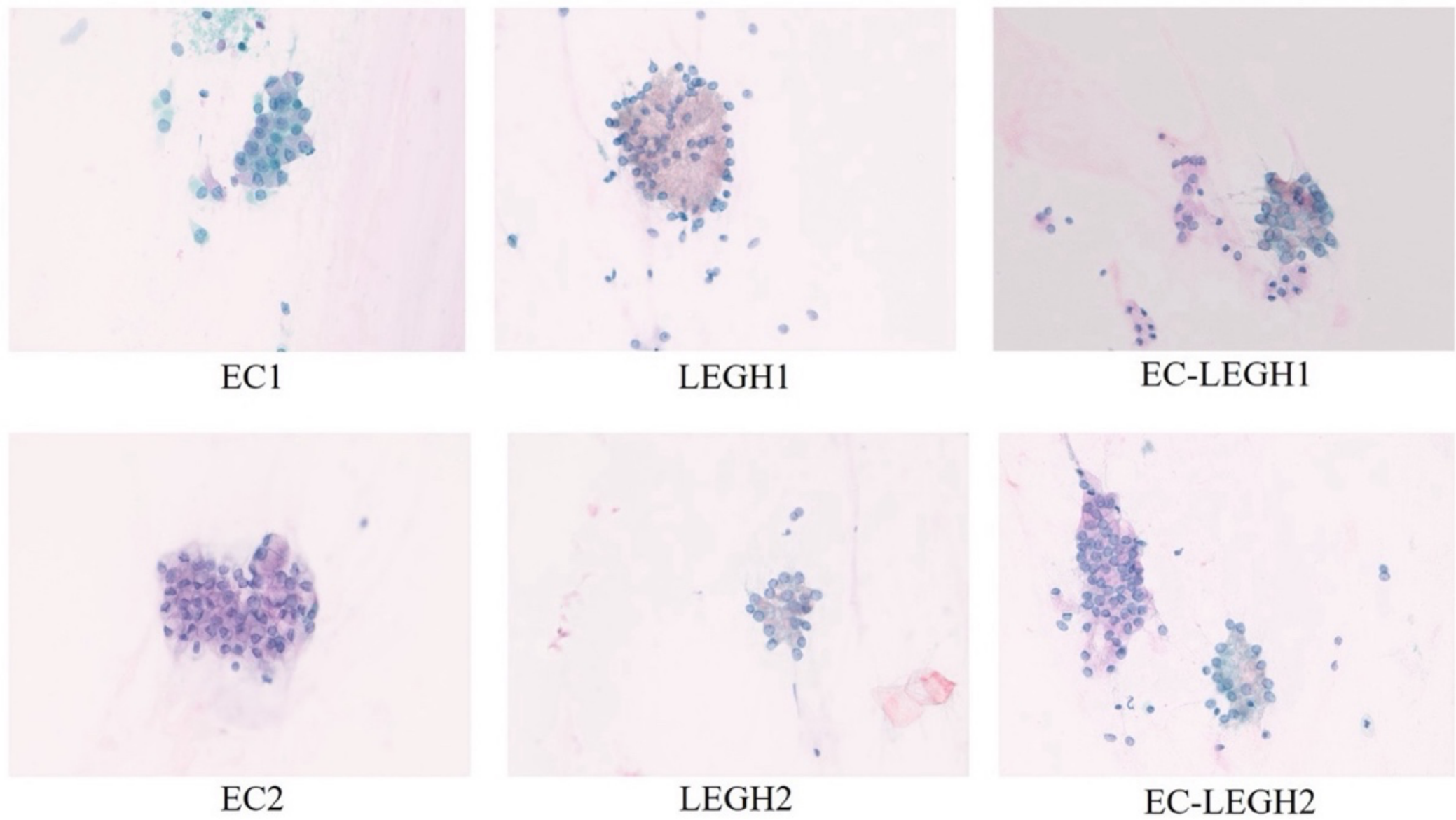

**Fig. 5** RGB images of cell samples. EC: (left), LEGH: (middle), mixed EC-LEGH: (right)

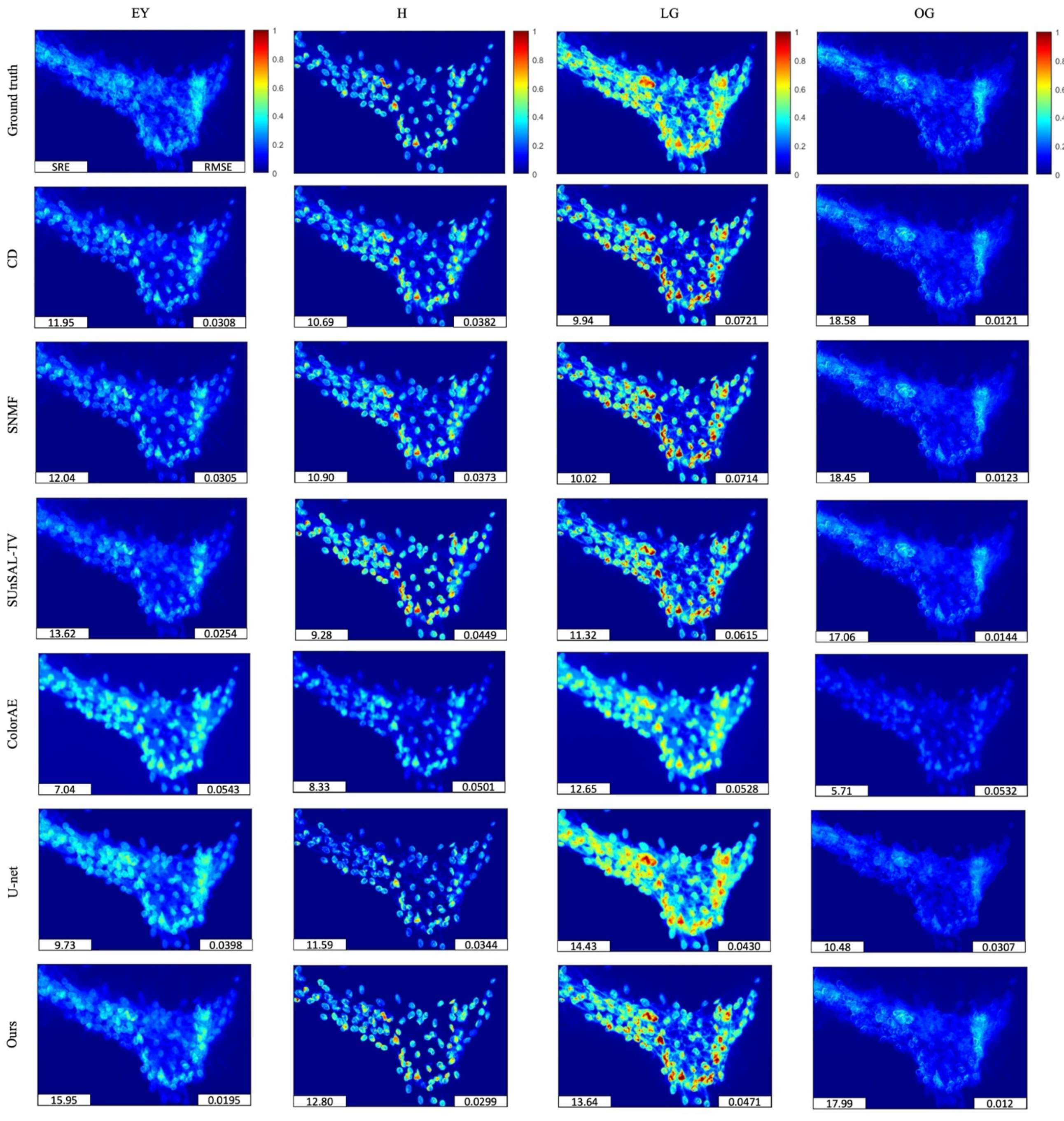

**Fig. 6** The ground truth stain abundances of EY, H, LG, and OG for one test image, and the estimated stain abundance maps obtained using modified CD, SNMF, SUnSAL-TV, ColorAE, U-Net, and the proposed method. Each abundance map includes two evaluation metrics: SRE (bottom left) and RMSE (bottom right), both computed separately for each dye

First, for the optimization-based methods (SNMF, SUnSAL-TV, and ours), we randomly selected six images from different slides for parameter tuning: two images containing EC cells, two with LEGH cells, and two containing both EC and LEGH cells, as illustrated in Fig. 5. We then computed the root mean square of six parameter sets to determine the tuned parameters and for SNMF ($\lambda = 5 \times 10^{-4}$), SUnSAL-TV ($\lambda = 3 \times 10^{-4}, \lambda_{TV} = 3 \times 10^{-3}$), and the proposed method ($\lambda = 2 \times 10^{-6}, \lambda_{TV} = 1 \times 10^{-3}$). We found that the optimal parameter values of our method were highly consistent across different images, indicating that a parameter set optimized for one image often performs well on others. In contrast, the optimal parameter sets of SUnSAL-TV varied substantially from image to image, suggesting that parameters effective for one image may not generalize well to others.

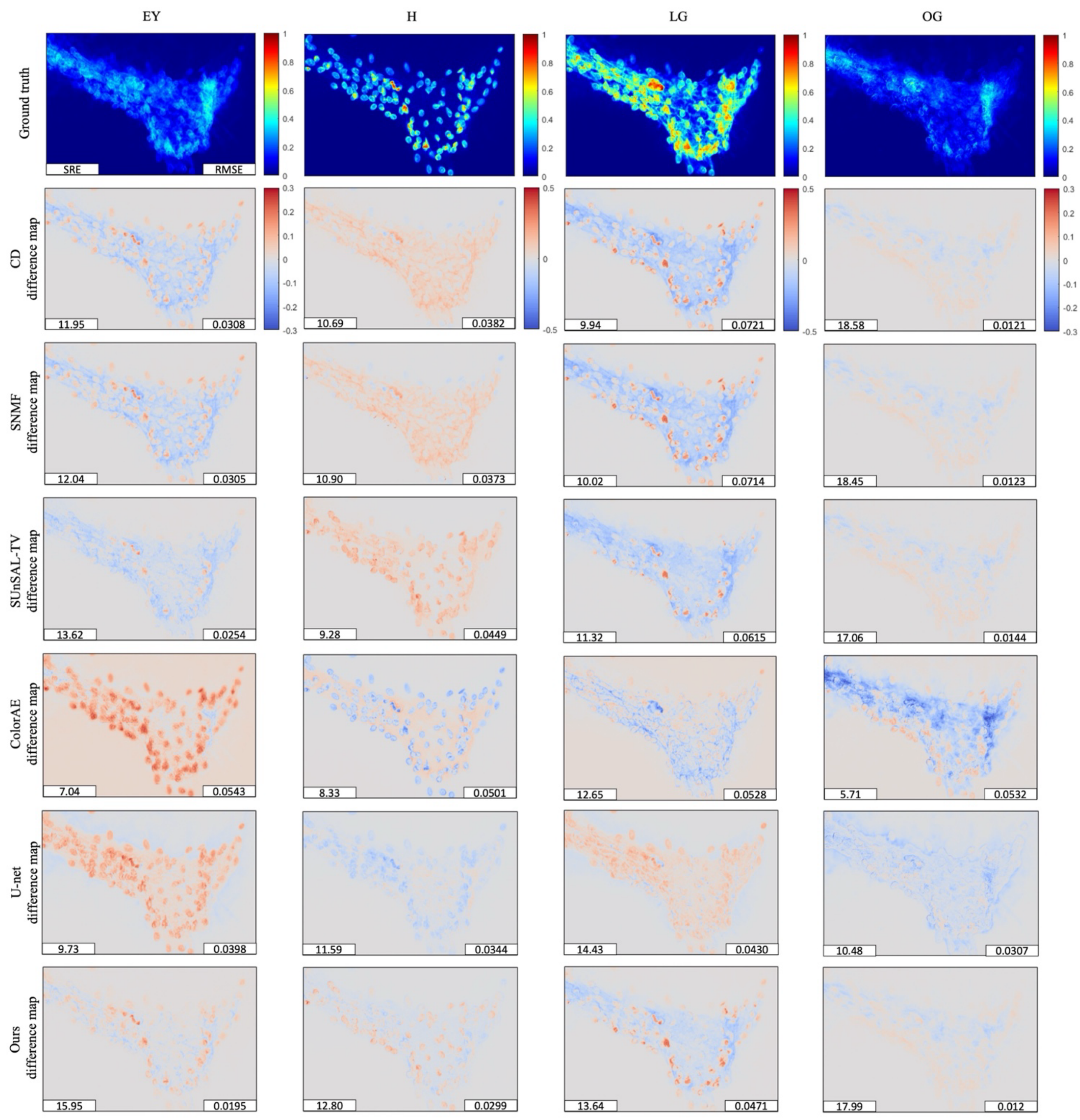

**Fig. 7** The difference maps for the same image, comparing the ground truth abundances of EY, H, LG, and OG with the results from modified CD, SNMF, SUnSAL-TV, ColorAE, U-Net, and the proposed method. Each difference map includes two evaluation metrics: SRE (bottom left) and RMSE (bottom right), both computed separately for each dye

Second, for the two deep learning-based methods, all RGB images in the training set were cropped into 256×256 pixel patches for training, with rotation and flipping used for data augmentation. For ColorAE, the reconstruction loss was computed as the mean squared error (MSE) between the reconstructed and input images on a pixel-wise basis, and an additional MSE abundance loss was calculated with respect to the MS stain unmixing results, with equal weighting assigned to the two losses. For U-Net, only the MSE abundance loss with respect to the MS stain unmixing results was used.

Then, we applied the parameter-tuned or trained methods to the test images. Fig. 6 presents the ground truth stain abundances of EY, H, LG, and OG for one test image, and the estimated stain abundance maps obtained using CD, SNMF, SUnSAL-TV, ColorAE, U-Net, and the proposed method. Fig. 7 shows the corresponding difference maps for the same image, comparing the ground truth abundances of EY, H, LG, and OG with the results from CD, SNMF, SUnSAL-TV, ColorAE, U-Net, and the proposed method. For each abundance map and difference map, except for the ground truth, two evaluation metrics, SRE and RMSE, are reported at the lower left and lower right corners, respectively. Qualitatively, the ground truth abundance maps show that all four dyes are visibly present in cell samples. For the traditional methods, the performance of stain unmixing improved as more regularizations were introduced. The results of our method, which incorporates the weighted nucleus sparsity constraint, were noticeably better than those of SUnSAL-TV, which employs the conventional l1 sparsity term. The difference maps further reveal that, among all methods, the proposed method yielded the smallest or second-smallest deviations from the ground truth across almost all dyes. Regarding the deep learning methods, ColorAE produced blurry abundance maps, giving the worst results for EY, H, and OG, but performing better than most traditional methods on LG. In contrast, U-Net generated sharper results and achieved the best performance on LG.

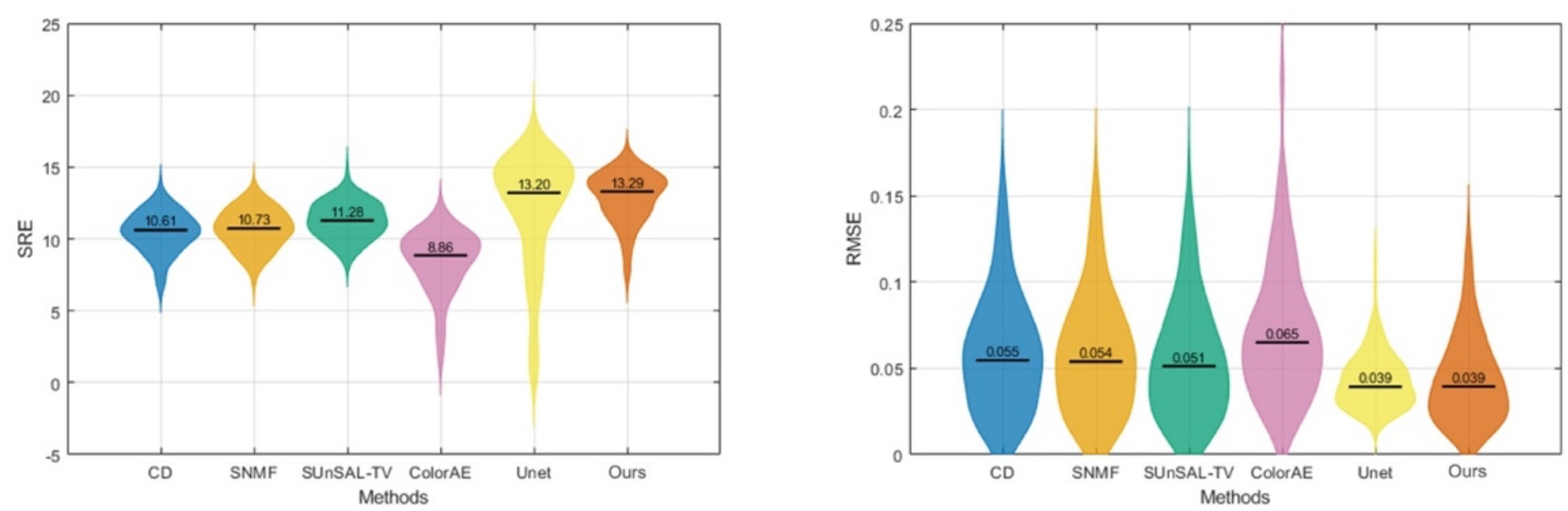

**Fig. 8** Violin plots of overall SRE (left) and RMSE (right) achieved by CD, SNMF, SUnSAL-TV, ColorAE, U-Net, and the proposed method on the test image patches. The black horizontal bars indicate the median values

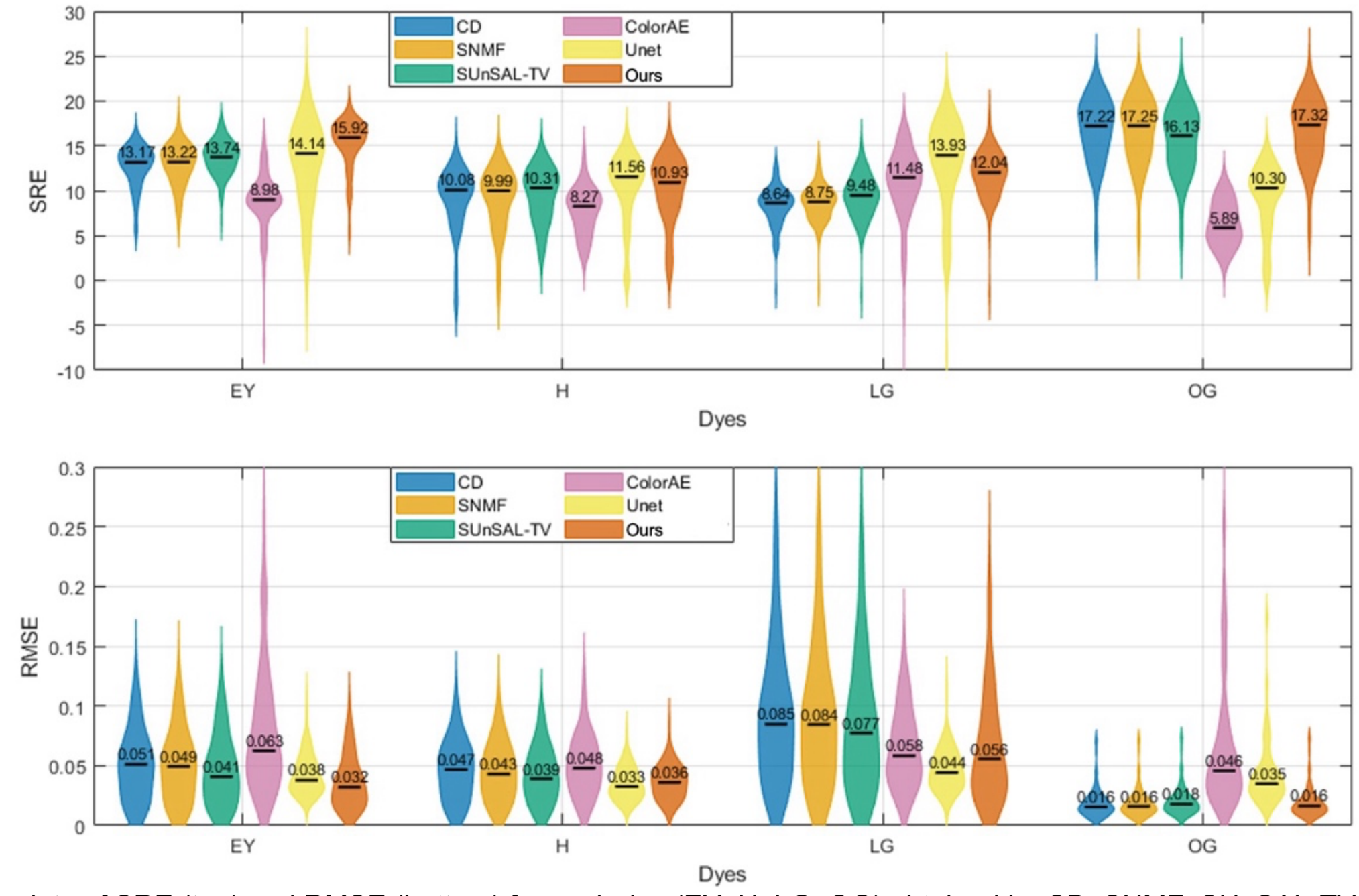

**Fig. 9** Violin plots of SRE (top) and RMSE (bottom) for each dye (EY, H, LG, OG) obtained by CD, SNMF, SUnSAL-TV, ColorAE, U-Net, and the proposed method on the test image patches. The black horizontal bars indicate the median values

Fig. 8 and Fig. 9 quantitatively compare the unmixing performance of CD, SNMF, SUnSAL-TV, ColorAE, U-Net, and the proposed method on the test image patches based on SRE and RMSE. In Fig. 8, which summarizes the overall performance across all dyes, our method achieved the highest median SRE (13.29 dB) and the lowest median RMSE (0.039), outperforming U-Net (13.20 dB and 0.039), SUnSAL-TV (11.28 dB and 0.051), SNMF (10.73 dB and 0.054), CD (10.61 dB and 0.055), and ColorAE (8.86 dB and 0.065). It can be observed that the results of U-Net and the proposed method were relatively close. Compared with U-Net, the SRE values of our method exhibited a tighter distribution, while the RMSE values showed a broader spread. This suggests that for images with large stain abundances (i.e., high signal energy), our method tended to yield larger RMSE values, whereas for images with small stain abundances (i.e., low signal energy), it achieved smaller RMSE values. In contrast, U-Net maintained relatively stable RMSE values regardless of stain abundance, but this led to smaller SRE estimates for images with low stain abundance.

A further analysis at the individual dye level shows that the proposed method outperformed all traditional methods in terms of both SRE and RMSE across all dyes, and it also surpassed all methods, including the deep learning ones, on EY and OG. The two deep learning methods exhibited a large performance gap: ColorAE delivered unsatisfactory results, whereas U-Net performed nearly as well as our method. Both deep learning methods performed well on LG, but their performance on OG was inferior to that of the traditional methods.

### *4.6 Parameter sensitivity and runtime*

In this section, we analyze the impact of regularization parameters in optimization methods and the runtimes of all methods. Fig. 10, using SNMF, SUnSAL-TV, and the proposed method as examples, illustrates the relationship between the SRE values obtained on a mixed EC-LEGH sample (Fig. 5 EC-LEGH1) and the parameters ($\lambda$ and $\lambda_{TV}$). From Fig. 10, it can be observed that our method achieved higher SRE values than the optimal results of the comparative methods over a wide range of parameters, making it easier to attain better unmixing results. Notably, the performance of our method did not peak when any parameter approached zero, suggesting that removing the corresponding regularization reduced effectiveness. Optimal performance was observed when $\lambda$ and $\lambda_{TV}$ approached $2 \times 10^{-6}$ and $1 \times 10^{-3}$, respectively. This suggests that the combined effect of weighted nucleus sparsity and TV regularization is crucial. In contrast, the l1 sparsity regularization in the comparative methods showed no significant improvement in SRE values across its parameter range, highlighting its ineffectiveness in Papanicolaou stain unmixing.

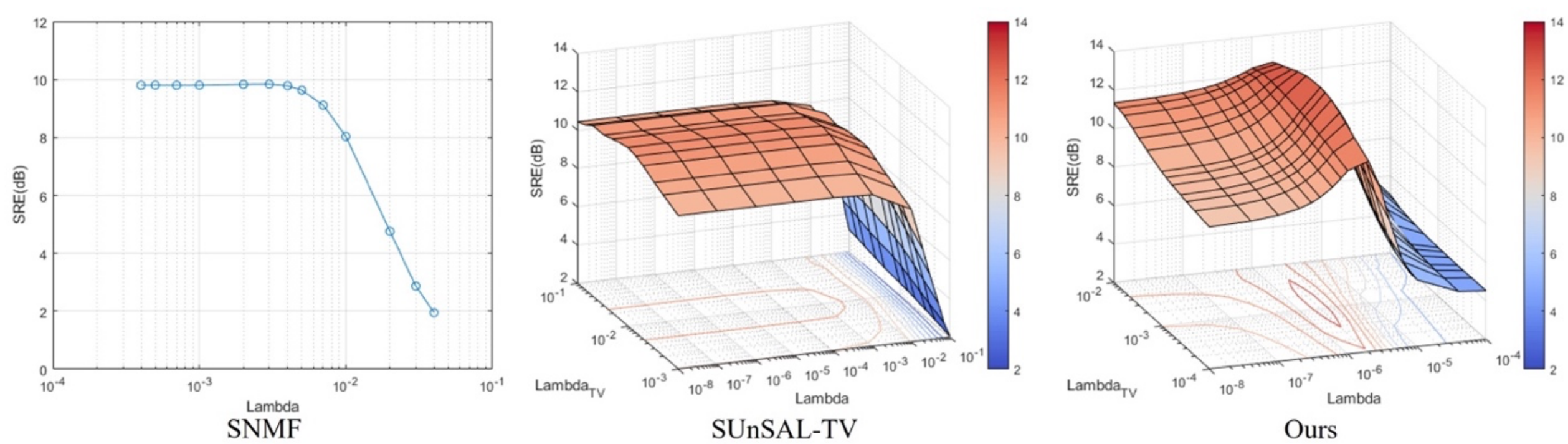

**Fig. 10** SRE as a function of regularization parameters for an EC-LEGH sample (Fig. 5 EC-LEGH1): $\lambda$ for SNMF (left), $\lambda$ and $\lambda_{TV}$ for SUnSAL-TV (middle), and for the proposed method (right)

**Table 3** Per-patch runtime over 10 random 256×256 Papanicolaou patches

| | CD | SNMF | SUnSAL-TV | ColorAE | U-Net | Ours |
|---|---|---|---|---|---|---|
| Time (s) | 0.0018 ± 0.0002 | 2.52 ± 0.37 | 14.56 ± 0.23 | 0.017 ± 0.004 | 0.031 ± 0.006 | 9.96 ± 0.82 |

[a] CD, SNMF, SUnSAL-TV, and the proposed method were executed on the CPU, while ColorAE and U-Net were executed on the GPU (only the time for a single forward propagation).

Then, we report runtimes for stain unmixing methods. All experiments were conducted on a desktop PC with an Intel Core i9-12900 CPU (2.40 GHz, 32 GB RAM) and an NVIDIA RTX A1000 GPU (8 GB). Modified CD, SNMF, SUnSAL-TV, and the proposed were executed on the CPU, while ColorAE and U-Net were executed on the GPU. We randomly selected ten 256×256 pixel Papanicolaou-stained patches and measured the time to complete one unmixing pass; Table 3 lists the mean ± standard deviation across methods. All optimization parameters were fixed to those in section 4.5, with a maximum of 1000 iterations. For the deep learning methods, the reported runtime reflects only a single forward propagation, excluding the training time. We observed that CD is the fastest baseline, as the other methods are optimization or deep learning methods. ColorAE and U-Net are faster than optimization-based methods, but they require training with substantial amounts of ground truth data. Among the optimization methods, incorporating TV increases the computational cost of SUnSAL-TV and our method; however, our method runs faster than SUnSAL-TV because our method typically converges after approximately 700 iterations. Specifically, our proposed method required 9.96 ± 0.82 s per 256×256 patch, with a peak memory usage of ~3.1

GB. It took ~10 hours to process the stained regions of a 40,000 × 40,000 WSI. The runtime can be further reduced through algorithmic optimization and parallel computation.

### *4.7 Cell classification based on stain unmixing*

To quantitatively analyze the differences in stain abundance between EC and LEGH cells, we extracted 248 EC mucin patches and 256 LEGH mucin patches, each measuring 5×5 pixels (Fig. 11), from the cytoplasmic regions of Papanicolaou-stained images corresponding to EC and LEGH cases, respectively. Then, we calculated the average stain abundance $\bar{\mathbf{x}}$ for each patch. Given that the stain abundance $\bar{\mathbf{x}}$ may correlate with the concentration and thickness of mucin in the cytoplasm, it is appropriate to compare the relative stain abundance $\bar{\mathbf{x}}_{\text{relative}}$, which is invariant to the thickness. It is calculated as follows:

$$\bar{\mathbf{x}}_{\text{relative},i} = \frac{\bar{\mathbf{x}}_i}{\sum_{i=1}^{4} \bar{\mathbf{x}}_i}. \tag{22}$$

where $\bar{\mathbf{x}}_i$ and $\bar{\mathbf{x}}_{\text{relative},i}$ are the average stain abundance and relative stain abundance of the $i$-th dye, respectively.

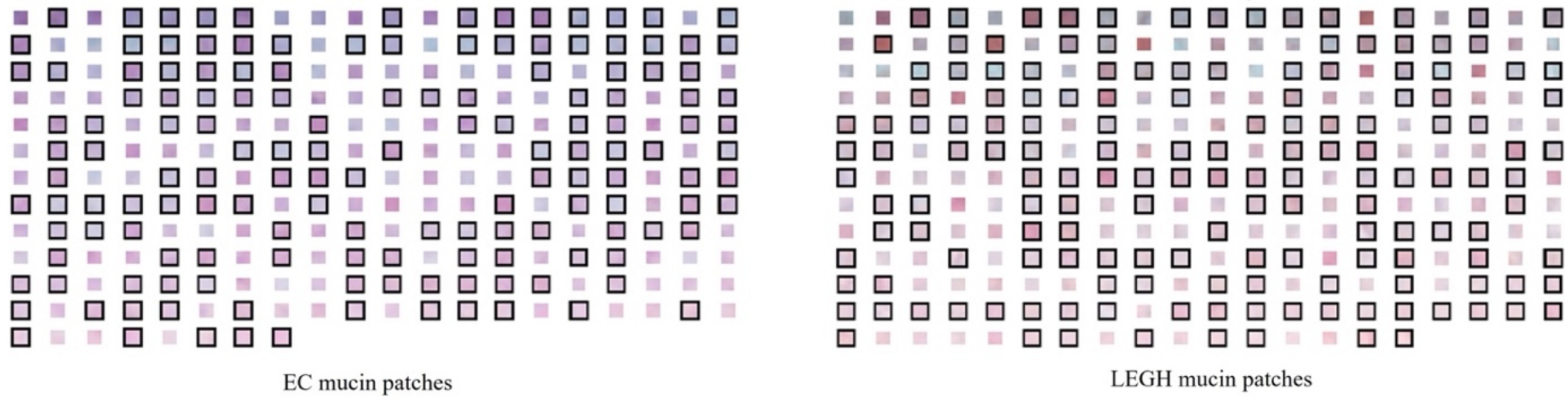

**Fig. 11** Patches of EC and LEGH mucin (patches in black boxes are the training data, and the rest are the test data)

The Mann-Whitney U-test, a nonparametric test, was used to determine if there were significant differences in relative stain abundance between LEGH and EC cells. The MS stain unmixing results were considered the ground truth. Fig. 12 presents box plots of the relative stain abundance for each dye derived from (a) MS stain unmixing and (b) the proposed RGB stain unmixing method, along with their p-values.

According to the results of the proposed method, significant differences are observed in EY, LG, and OG relative abundances ($p < 0.001$). In the MS results, significant differences exist in the relative abundances of EY, H, and OG ($p < 0.001$). The discrepancies between methods reside in the H and LG channels. The larger p-value for H produced by our method is attributable to the spectral limitations of RGB images. The significant LG difference detected by the RGB-based method—but not by the MS analysis—may stem from the RGB model's assumption of a linear relationship between stain abundance and RGB absorbance. Despite these discrepancies, both subfigures consistently show that LEGH samples have a noticeably lower relative abundance of EY and a higher abundance of OG compared to EC samples. Therefore, the relative stain abundances of EY and OG obtained through stain unmixing can be effectively used to distinguish between EC and LEGH cell mucin.

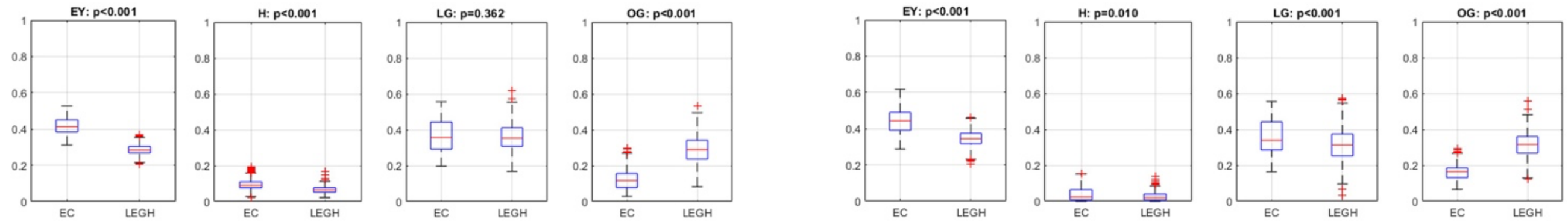

**Fig. 12** Box plot of the relative stain abundance of each dye with each p-value using the Mann-Whitney U-test: MS stain unmixing (left), and the proposed RGB stain unmixing method (right)

To further assess the utility of stain abundance quantification in cytological diagnosis, we employed relative stain abundances derived from our proposed unmixing method to classify mucin in EC and LEGH cells. To maintain high interpretability in the classification process, we selected linear discriminant analysis (LDA) as our classifier. For comparative analysis, we also trained LDA models using RGB intensities, OD values, CIELAB values, and relative stain abundances obtained from MS unmixing, respectively. The following feature sets were used to train the LDA classifiers:

- RGB intensities: The model used the raw red, green, and blue channel intensities as input features.
- OD values: The model used the OD values computed from the red, green, and blue channels.
- CIELAB values: The model used the L*, a*, and b* components of the CIELAB color space.

- MS-based stain abundances: The model used the relative stain abundances (EY, H, and OG) obtained from MS stain unmixing.
- Proposed RGB-based stain abundances: The model used the relative stain abundances of EY and OG derived from the proposed RGB stain unmixing method.

From the patches illustrated in Fig. 11, we randomly selected 148 EC and 156 LEGH mucin patches as the training set, with the remaining 100 EC and 100 LEGH patches reserved for testing. The resulting LDA score density and classification plots for each LDA model are shown in Fig. 13, where EC patches are marked with circles, LEGH patches with squares, and misclassified patches with crosses. The decision boundary is depicted either by a black line or a yellow plane. Gray-shaded areas in the LDA score density plots indicate where misclassification occurred.

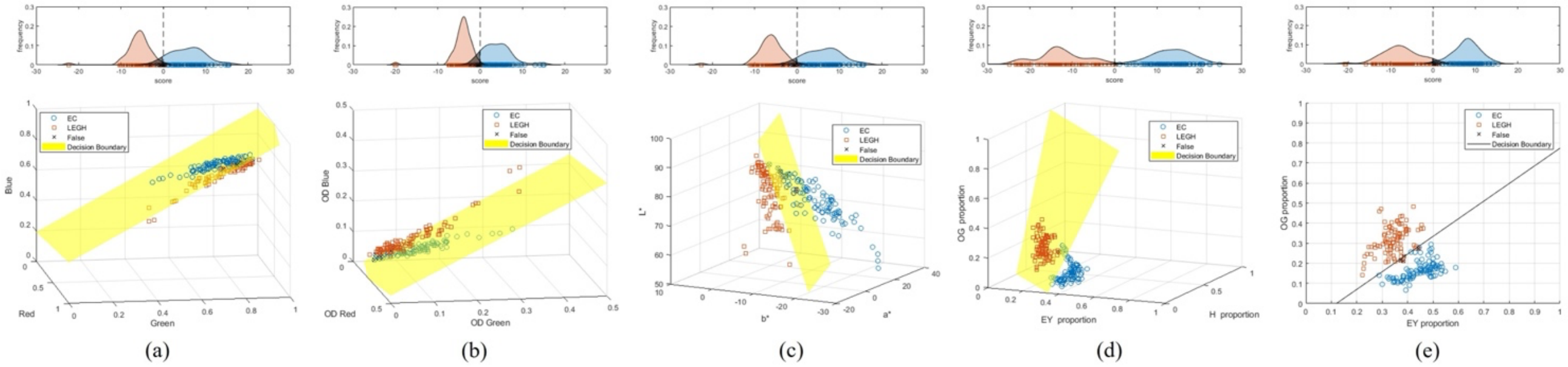

**Fig. 13** LDA score density plots (top row) and corresponding classification results (bottom row) using different feature sets: (a) RGB intensities, (b) OD values, (c) CIELAB values, (d) relative stain abundances from MS unmixing, and (e) relative stain abundances from the proposed RGB unmixing method

Table 4 presents the four key quality metrics of the classification results—accuracy, precision, recall, and F1 score—which together provide a comprehensive evaluation of each LDA model's performance. The linear discriminant function $D(\boldsymbol{x})$ of the proposed classification method was derived as follows:

$$D(\boldsymbol{x}) = 57.96EY - 65.84OG - 6.99. \tag{23}$$

The patch was classified into LEGH if $D(\boldsymbol{x}) \geq 0$ and EC if $D(\boldsymbol{x}) < 0$.

The numerical results indicate that the LDA classifier based on the proposed RGB unmixing method achieved outcomes closest to those based on MS stain unmixing results, outperforming classifiers based on RGB intensity, OD, and CIELAB value. As illustrated by Fig. 11, it is difficult to visually distinguish between EC and LEGH cells based on color differences. The separations of EC and LEGH patches based on RGB intensity, OD, and CIELAB value show less clarity, as reflected by larger overlapping areas in the LDA score density plots. However, the distinction between EC and LEGH patches is more pronounced when using relative stain abundances derived from MS or RGB stain unmixing. The 2D classification plot reveals that the yellow hue of LEGH mucin corresponds to a higher relative abundance of OG and a lower relative abundance of EY. These results indicate that the proposed RGB stain unmixing method can effectively quantify dye amounts and that employing relative stain abundances as features enhances both the accuracy and explainability of distinguishing EC from LEGH samples.

**Table 4** Four quality measures of LDA models

| LDA model | Accuracy | Precision | Recall | F1 score |
|---|---|---|---|---|
| RGB intensity | 0.965 | 0.943 | 0.990 | 0.966 |
| OD | 0.945 | 0.908 | 0.990 | 0.947 |
| CIELAB value | 0.950 | 0.925 | 0.980 | 0.952 |
| Relative stain abundances from MS unmixing | 0.995 | 1.000 | 0.990 | 0.995 |
| Relative stain abundances from the proposed RGB unmixing | 0.980 | 1.000 | 0.960 | 0.980 |

## 5 Discussion

This paper introduces the first training-free stain unmixing method for RGB images of Papanicolaou-stained specimens. Utilizing the prior knowledge that H typically concentrates in the nuclei rather than the cytoplasm, we propose a novel constraint term, weighted nucleus sparsity regularization. This term is combined with nonnegativity and TV regularization to estimate effectively the stain abundances for four dyes (EY, H, LG, and OG) from three-channel RGB images by solving an optimization problem. In addition, our method treats images prepared under consistent experimental and imaging conditions as an individual domain, including staining reagents, protocols, and scanner settings, within which all images share a common stain matrix. This allows for direct comparison of stain abundances across slides in the same domain without normalization. In contrast, WSI-specific stain unmixing methods treat each WSI as an independent domain, requiring appropriate normalization when comparing across different WSIs.

In our experiments, the stain matrix was predetermined using single-stain slides, and the stain unmixing results from MS images [8] were regarded as the ground truth. The quantitative results (Sections 4.5–4.6) indicate that our proposed method achieved superior stain unmixing performance compared with both traditional and deep learning methods. The proposed weighted nucleus sparsity constraint effectively mitigated the overestimation of H in non-nuclear regions, thereby improving the estimation accuracy for all four dyes. However, the conventional sparsity term, assuming dyes are sparse (i.e., only a small number of dyes exist simultaneously) in the image, is not suitable for Papanicolaou-stained images. Among the deep learning methods, U-Net outperformed ColorAE, suggesting that skip connections effectively maintain spatial structural information, whereas the combined use of reconstruction and abundance losses in ColorAE may have introduced additional errors through the reconstruction term. Although the proposed method was slower than deep learning approaches when unmixing, it is scalable to arbitrary input sizes and avoids the need for extensive ground truth training data, which is often difficult to obtain in practice.

The quantitative results also demonstrated the utility of our stain unmixing method in downstream diagnostic tasks. By analyzing the estimated stain abundances, we confirmed statistically significant differences in the relative stain abundances of EY and OG between the cytoplasmic mucins of EC and LEGH cells. Furthermore, in the classification of mucin patches from EC and LEGH cells, our method achieved an accuracy of 98.0%, closely approaching the performance of MS-based stain unmixing (99.5%) and outperforming classifiers based on RGB, OD, and CIELAB features. In our experiments, we did not report cell classification results based on other stain unmixing methods for two reasons. First, for cell classification tasks, accurate estimation of relative stain abundances requires minimizing the error-to-ground-truth ratio, in which case the SRE metric is more critical than RMSE. Second, the differences in stain abundances between EC and LEGH cells are primarily reflected in EY and OG, for which our proposed method achieved the best unmixing performance among all methods. Therefore, it can be inferred that our method offers greater robustness in accurately classifying EC and LEGH cells compared with other approaches. These findings suggest that our method not only enhances the interpretability of diagnosis but also improves classification accuracy.

Recent studies have demonstrated that utilizing separated stain channels in H&E or IHC images can improve cell segmentation and classification [25,31,42,43]. As digital pathology continues to grow, stain unmixing is expected to play an increasingly significant role in enhancing the accuracy and reliability of computational cytology.

Nevertheless, several limitations of the current study merit discussion. First, although our method is training-free, it requires prior measurement of single-stain reference samples, which may not always be available in practice. Second, our method requires a separate stain matrix for each domain (e.g., Papanicolaou slides prepared in different institutions). How to perform cross-domain comparison of stain abundances remains to be investigated. Besides, although we provide a brief robustness analysis across different RGB acquisition settings, all images in this study were collected at a single institution; we did not evaluate variability arising from staining protocols, reagents, technicians, or environmental conditions. Third, the estimation of LG by the proposed method was less accurate than that of U-Net, because we observed that the RGB absorbance of LG does not exhibit a strictly linear relationship with its abundance, leading to errors when a linear unmixing model is applied. The nonlinearity should be addressed in future work. Fourth, because BY staining was negligible in our dataset, we used a four-dye model. The current implementation has not been validated on BY-rich slides and may require a five-dye unmixing or adaptive component selection in such cases.

Future work may focus on estimating the stain matrix for a set of WSIs. For instance, for slides processed within the same laboratory and batch, a domain-specific stain matrix can be estimated from a small set of representative Papanicolaou slides and then applied to the remainder. Besides, it is expected to incorporate larger, more diverse datasets to enable a systematic and comprehensive evaluation of the proposed approach's cross-domain generalizability. In addition, we plan to leverage external datasets to expand downstream tasks driven by stain abundance, e.g., cell segmentation, cell classification, and cancer subtyping, to further demonstrate the practical utility of stain unmixing in digital pathology. Finally, investigating the practical deployment of the proposed method within real cytopathology workflows will be a critical step toward clinical translation.

## 6 Conclusion

To date, limited research has specifically addressed stain unmixing in Papanicolaou-stained images. We have introduced and rigorously validated the first RGB-based Papanicolaou stain unmixing method. The proposed approach is training-free and integrates nonnegativity, TV smoothness, and a weighted nucleus sparsity regularization within a convex optimization framework. In qualitative and quantitative experiments, the proposed method achieved the most accurate stain quantification results compared with the MS unmixing results, without requiring extensive manual annotation or ground truth data. The estimated abundance maps yield clinically meaningful features that distinguish LEGH cells from normal endocervical cells with high precision. By turning subjective color impressions into quantitative stain abundances, the technique offers an objective, resource-efficient foundation for cervical cancer screening and paves the way for broader adoption of stain unmixing in digital pathology, especially when the number of dyes exceeds the available image channels.

## Declarations

**Funding**: This work was supported by JST SPRING, Japan Grant Number JPMJSP2106 and JPMJSP2180, and Engineering Research Grants, Mizuho Foundation for the Promotion of Sciences. This paper was partly based on the results from project JPNP20006, subsidized by the New Energy and Industrial Technology Development Organization (NEDO).

**Conflict of Interest**: The authors declare that they have no conflict of interest.

**Ethics approval**: This study was performed in line with the principles of the Declaration of Helsinki. Approval was granted by the Medical Ethics Committee of Shinshu University (Nov. 6, 2024, No. 5470).

**Consent to participate**: Informed consent was obtained from all individual participants included in the study.

**Consent to publish**: Patients signed informed consent regarding publishing their data and photographs.

**Code availability**: The implementation of the proposed Papanicolaou stain unmixing method is publicly available at https://github.com/ChrisG0ng/SUnWNS-TV.

**Authorship contribution**: Nanxin Gong: Conceptualization, Investigation, Methodology, Software, Formal analysis, Visualization, Writing – original draft. Saori Takeyama: Investigation, Data curation, Software, Writing – review. Masahiro Yamaguchi: Conceptualization, Supervision, Project administration, Funding acquisition, Writing – review. Takumi Urata: Data curation, Validation. Fumikazu Kimura: Conceptualization, Validation, Resources, Supervision, Writing – review. Keiko Ishii: Validation, Resources, Writing – review.

## Author biography

Nanxin Gong is a Ph.D. student in the Department of Information and Communications Engineering, School of Engineering, Institute of Science Tokyo.

Saori Takeyama is an Assistant Professor in the Department of Information and Communications Engineering, School of Engineering, Institute of Science Tokyo.

Masahiro Yamaguchi is a professor in the Department of Information and Communications Engineering, School of Engineering, Institute of Science Tokyo.

Takumi Urata is a Ph.D. student in the Department of Information and Communications Engineering, School of Engineering, Institute of Science Tokyo.

Fumikazu Kimura is a Junior Associate Professor in the Department of Biomedical Laboratory Sciences, School of Health Sciences, Shinshu University.

Keiko Ishii received a Medical Doctor and Ph.D. from Shinshu University School of Medicine. She is working as a Pathologist at the Division of Diagnostic Pathology, Okaya City Hospital.